# Spatial inhomogeneity and chiral symmetry of the lattice incommensurate supermodulation in high temperature superconductor $Bi_2Sr_2CaCu_2O_{8+y}$

Nicola Poccia[1], Gaetano Campi[2], Michela Fratini[1], Alessandro Ricci[1], Naurang L. Saini[1], Antonio Bianconi[1*]

[1]Department of Physics, Sapienza University of Rome, P. le A. Moro 2, 00185 Roma, Italy
[2]Institute of Crystallography, CNR, via Salaria Km 29.300, Monterotondo Stazione, Roma, I-00016, Italy

**Using scanning micro X-ray diffraction we report a mixed real and reciprocal space visualization of the spatial heterogeneity of the lattice incommensurate supermodulation in single crystal of $Bi_2Sr_2CaCu_2O_{8+y}$ with $T_c$=84 K. The mapping shows an amplitude distribution of the supermodulation with large lattice fluctuations at microscale with about 50% amplitude variation. The angular distribution of the supermodulation amplitude in the a-b plane shows a lattice chiral symmetry, forming a left-handed oriented striped pattern. The spatial correlation of the supermodulation is well described by a compressed exponential with an exponent of 1.5±0.3 and a correlation length of about 50 $\mu$m, showing the intrinsic lattice disorder in high temperature superconductors.**

*E-mail: antonio.bianconi@uniroma1.it

One of the emerging common features of high temperature superconductors (HTSs) is the intrinsic lattice inhomogeneity [1-4] but it remains practically unknown because of the lack of methods probing both the k-space and the real space from nano-scale to micron-scale. The intrinsic lattice inhomogeneity is related with the co-existence of two or more electronic components showing a nanoscale phase separation in the normal phase [3-4]. In fact two strongly correlated electronic components at the Fermi level are expected to exhibit a phase separation between two electronic phases with different charge density [5]. The low density phase is close to the Wigner localization limit in the normal phase and below $T_c$ forms a Bose-like condensate or a bipolaronic condensate in presence of electron-lattice interaction [6,7]. The high density phase forms a BCS condensate below $T_c$, therefore the HTS phase appears to be at the BCS-Bose crossover in a multi-gap superconductor [6,7].

The striped superconductor showing nanoscale phase separation of striped domains with different gaps has been called the "superstripes" scenario characterized at lager scales by a complex landscape of striped patches [8].

In cuprates there is agreement on the experimental fact that the itinerant carriers induced by doping are injected in the oxygen 2p orbitals [9] and that at the Fermi level there are two main electronic components [10]:

1) the nodal particles with higher Fermi velocity with the wavevector on the Fermi surface arcs around the $(\pi,\pi)$ direction of the reciprocal k-space, i.e., moving nearly at 45° from the Cu-O bond direction. These states have molecular $b_1$ orbital symmetry made of atomic oxygen $2p_{x,y}$ hybridized with pure Cu $3d_{x^2-y^2}$ orbital.

2) the antinodal particles with lower Fermi velocity with Fermi wavevector in the $(0,\pi),(\pi,0)$ directions of the reciprocal k-space. In the normal phase they show a pseudo gap below T*. These particules have a strong coupling to the $B_{1g}$ phonon [11]. This mode involves the $Q_2$ rhombic distortion of the $CuO_4$ platelet [12]. These particles run in the real space around the Cu-O bond direction and are made of molecular orbital with $a_1$ symmetry made of oxygen $2p_{x,y}$ orbital hybridized with mixed Cu $3d_{3z^2-r^2}$ and $3d_{x^2-y^2}$ atomic orbital. Their orbital symmetry has been determined by Cu $L_{2,3}$ polarized XANES and the associated instantaneous pair distribution function of the Cu-O pairs by polarized EXAFS and they have been associated with pseudo Jahn Teller polarons [13].

The phase separation in cuprates with the formation of lattice domains with different charge densities has been clearly observed in the case of mobile dopants, oxygen interstitials in oxygenated $La_2CuO_{4+y}$ both in the underdoped [14] and in the optimum doping [15,16] regime. In this case the domain growth and the density of mobile oxygen interstitials are expected to track the hole density in the $CuO_2$ plane.

A second common feature of all HTSs is that their lattice architecture is a "heterostructure at atomic limit" where atomic "*layers*" of a superconducting material are intercalated by block "*layers*" [17]. In these superlattices the lattice misfit between the different layers both in cuprates [18-22], pnictides [23] and chalcogenides [24,25] plays a key role as the third variable in the phase diagram of the superconductors.

Most of the studies on electronic inhomogeneity have been focused on the system $Bi_2Sr_2CaCu_2O_{8+y}$ (Bi2212). In this system the misfit strain [18-22] between the block layers ($Bi_2O_2$-$Sr_2O_2$) and the bilayers ($CuO_2$-Ca-$CuO_2$) induces a lattice corrugation of both the BiO planes [26-29] and the $CuO_2$ planes [30,31] appearing as a lattice supermodulation in x-ray diffraction that form stripes of distorted and undistorted lattice witan electronic structure reconstruction with multi bands and broken Fermi surface arcs [31-34]. This lattice supermodulation has been recently associated with the modulation of the superconducting gaps [35] revealing the effect of this specific lattice modulation on the superconducting $CuO_2$ layers [30,31].

In this work we have investigated the spatial mapping of the variation of the amplitude of the superstructure lattice modulation with the specific wavevector 0.21b*, 0.5c* in Bi2212 [26-31] looking for multiscale phase separation on the micron scale [38,39]. The nanoscale electronic and topological inhomogeneity in Bi2212 has been widely investigated by scanning tunneling microscopy/spectroscopy (STM/STS) [35, 40-43]. The Fermi surface has been measured by techniques like ARPES [10,44]. The k-space information obtained from these methods is spatially averaged, and it is generally difficult to quantitatively reconcile these results with the nanoscale structure. Here, we have attempted to overcome this limit and used scanning micro X-ray diffraction (μXRD). The results provide a mixed real and reciprocal space visualization of the spatial heterogeneity of the lattice incommensurate supermodulation in the Bi2212 system, showing quenched disorder and translational and rotational broken symmetry.

The μXRD experiment at the beam-line (ID13) of the ESRF takes advantage of the novel focusing method of the 12KeV x-ray synchrotron radiation beam emitted by a 18 mm period undulator of ESRF 6.03 GeV storage ring, operated in multibunch mode with a current of 200 mA. The beam-line uses two monochromators positioned in series; the first is a liquid $N_2$ cooled Si-111 double crystal or Si-111 (bounce); the second, is a channel cut monochromator employing a single liquid nitrogen cooled Si crystal. The optics of the microfocus beamline includes compound refractive lenses, Kirkpatrick Baez (KB) mirrors and crossed Fresnel zone plates focusing the beam to a spot of 1 micron in diameter. The interaction volume with the crystal of the 1 micron beam is $1 \times 1 \times 20 \mu m^3$ .he detector of x-ray diffraction images is a high resolution CCD camera (Mar CCD) with point spread function 0.1 mm, 130 mm entrance window, 16 bit readout placed at a distance of about 90 mm from the sample. The measurements were made on a well-characterized Bi2212 single crystal of 100 x 130 μm$^2$ surface and 80 μm thickness, grown by the traveling-solvent-floating-zone method (TSFZ). The growth velocity was adopted to be 0.5mm/h, and the growth atmosphere was a mixed gas flow of $O_2$ (20%) and Ar (80%). Structural analysis was performed using a 4-axis X-ray diffraction, and the basic structure of the samples was found to be orthorhombic symmetry with lattice parameter of a≈542 pm , b≈547 pm, c=3070 pm, α and γ ≈90.0°, β ≈92° with a large orthorhombic distortion in the range 0.5-1.1 % in agreement with other overdoped crystals grown by TSFZ technique [45,46]. The investigated samples are overdoped with an oxygen content y determined to be around 0.26±0.01, corresponding to a hole density of about 0.21 holes per Cu site [47]. The crystal was mounted on an XY mechanical translator for scanning parallel to the crystallographic (a,b) plane. The experimental set-up in x-ray reflection mode allows the xy-translation of the sample with 4.97 μm steps in the x and y directions .

A typical diffraction pattern of synchrotron X-ray reflections due to satellite peaks near the main crystal reflection (0,0,20) is shown in Fig. 1a. There are two distinct sets of incommensurate superstructures. The first is a three-dimensional long-range order (3D-LRO) modulation, characterized by a very sharp and resolution limited peak , the second is a two-dimensional (2D) short-range order (SRO) modulation, characterized by a diffuse profile in the c-axis direction due to 2D domain walls separating the 3D domains of 3D-LRO. For each Bragg refection the diffraction vector H can be written as $H = ha^* + kb^* + 1c^* + mq_s [q_s = \beta b^* + (1/\gamma)c^*, \gamma = 1]$.

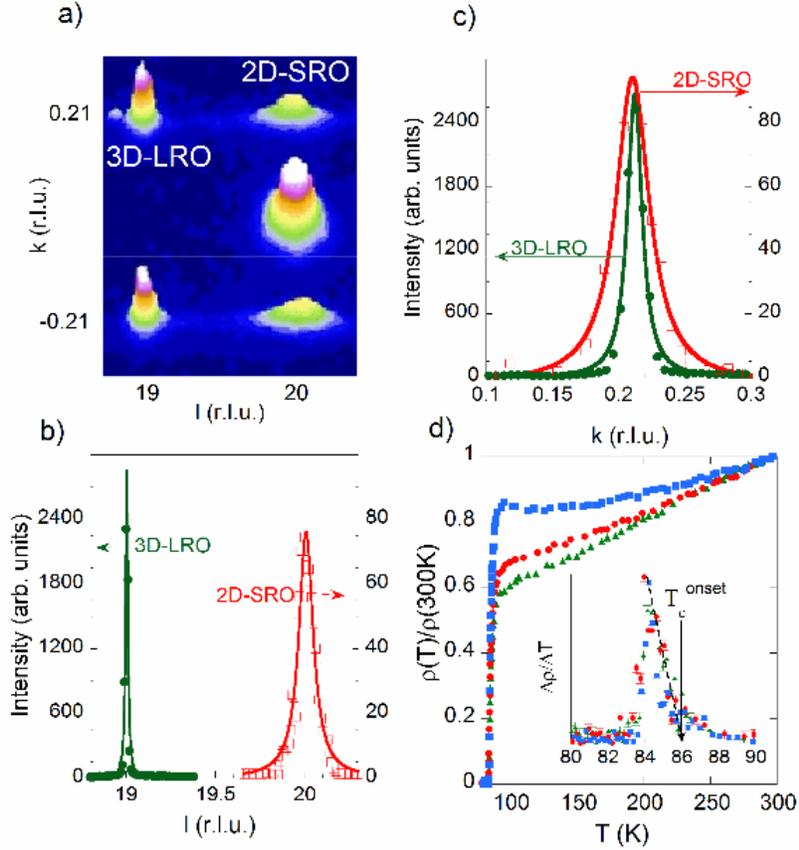

**Figure 1: a)** X-ray diffraction pattern in the (k,l) plane of Bi2212 recorded by a CCD detector showing the 3D-LRO and 2D-SRO satellites near the $(0,0,20)$ reflection; The **b)** and **c)** panels show the profiles of the x-ray reflections due to the 2D-SRO (empty rectangles) and 3D-LRO (filled circles) satellites along c* and b* directions where l and k are measured in the reciprocal lattice units (r.l.u) respectively. **d)** Temperature dependent resistivity of Bi2212 superconducting crystals used in this work (green, red and blue curves represent resistivity in the a, b and c axis respectively). The insert to the panel shows the resistivity derivatives around the $T_c$. The black arrow shows the $T_c$ onset of the superconducting transition.

The $q_s$ vector, called superstructure vector, is the wavevector of the modulation, defined as the linear combination of the base vectors of the three dimensional reciprocal lattice $q_s = \alpha a^* + \beta b^* + \gamma c^*$. Fig 1b and Fig. 1c show the profiles of the x-ray scattering reflections $(h, 4-n\times\tau, l)$ for $\tau = 0.21$. The experimental diffraction coherence lengths are $\xi_c^{3D} = 1/\Delta L = 944 \pm 2\ nm;\ \xi_b^{3D} = 1/\Delta k = 190 \pm 2\ nm$ and $\xi_c^{2D} = 1/\Delta L = 146 \pm 5\ nm;\ \xi_b^{2D} = 1/\Delta K = 82 \pm 5\ nm$, where Δl, Δk are the FWHM of the diffraction reflection profiles, respectively for the 3D-LRO and 2D-SRO, distinguishing the long-range from the short-range order.

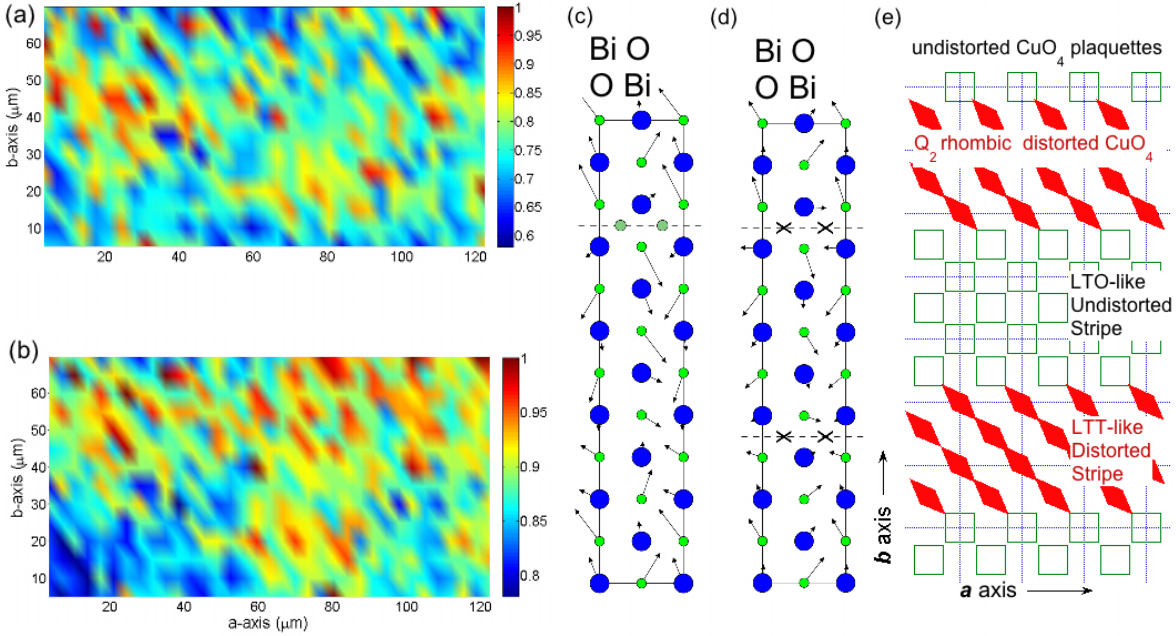

**Figure 2:|** (color online) Panels **a** and **b**: The position dependent intensities of the superstructure 3D-LRO (panel **a)**) and the superstructure 2D-SRO (panel **b)**) superstructure satellite reflections in Bi2212. The crystallographic **a**-axis is along the horizontal direction, and the **b**-axis is along the vertical direction. The very dark gray (blue color) indicates weak supermodulation amplitude, the light gray and dark gray (yellow and red) show the regions of high XRD superstructure reflections.. Panel (**c**) and panel (**d**) (from reference 36,37**:** Atomic displacements in the BiO layer projected into the x-y plane from DFT bulk calculation (36,37) (panel c) and the. experimental data by diffraction (see, e.g. Ref. 28); dashed lines are mirror planes, and X's mark points of stability found for oxygen interstitials by DFT (Panel **d**) Panel **e** shows the topological distribution of lattice stripes in the $CuO_2$ plane of BI2212 plane probed by Cu K-edge resonant x-ray diffraction [30] and by Cu K-edge EXAFS [31] and The empty squares (filled rhombs) indicate undistorted (distorted) $CuO_4$ plaquettes in the undistorted (distorted) stripes. The rhombs due to atomic $Q_2$ local lattice distortion of the square $CuO_4$ plaquette could order in the right-handed or the left-handed direction breaking the rotational symmetry of the lattice. The figure shows the left handed orientation of the $Q_2$ rhombic distortions with the direction of the Cu-O bond elongation of the distorted $CuO_4$ plaquette at 45° from the direction of the supermodulation parallel to the vertical direction.

The x–y position dependence of the integrated satellite peak intensity (Fig. 2a,b) shows that the coexisting 2D-SRO and 3D-LRO structural modulations are quite inhomogeneous on the micrometer scale. The position dependence of the 3D-LRO and 2D-SRO satellites is shown in Fig. 2 (panels **a** and **b** respectively). The color scales from dark blue (minimum) to the dark red (maximum). The intensity (satellite superstructure intensity integrated over a square sub-areas of the CCD detector) is normalized with respect to the background intensity ($I_o$) of the tail of the main crystalline reflections. The ordered patches of 3D-LRO and 2D-SRO are oriented along one specific direction (left direction), 45° with respect to the vertical direction (superlattice direction) implying two different diagonal directions (left and right) in the real space. This result provides a clear evidence of the spontaneous breaking of the rotational crystal symmetry in the plane.

It was known that the atomic displacements in the BiO layers [26-29] run in the diagonal direction confirmed by ab-initio calculations [36,37]. Cu-K-edge resonant x-ray diffraction at the Cu K-edge have shown that the $CuO_2$ plane is decorated by stripes of distorted and undistorted lattice in Bi2212 [30] and the polarized extended x-ray absorption fine structure (EXAFS) data [48,49] have shown that anomalous Cu-O-Cu bonds (45° from the superlattice direction) appear along only one of the two (left handed or right handed) diagonal directions as shown in Fig. 2c. The broken rotational symmetry has also been observed later by STM experiments [43].

The present evidence of left-handed micro-scale inhomogeneity of the lattice supermodulation intensity fluctuations provides evidence for broken lattice symmetry on a scale three order of magnitude larger than seen before. This result confirms the proposed intrinsic multiscale structural complexity driven from nanoscale to microscale by elastic fields [38,39] in a popular model cuprate system.

In Fig. 3 and Fig. 4 we quantify the observed inhomogeneity of the 3D-LRO and 2D-SRO superstructures, by plotting the distribution of intensities and the distance-dependent intensity correlations. Fig. 3 shows that the mean values of the amplitude of the 3D-LRO and 2D-SRO are different, the deviation from the average for both supermodulations is around 50-60%. Also, over a small 50 μm area the amplitude of the micro-XRD reflection fluctuations are of the order of 30%. The large disorder of the amplitude of supermodulation can be assigned to the spatial inhomogeneity due to space variable stoichiometric that strongly deviates from the nominal composition.

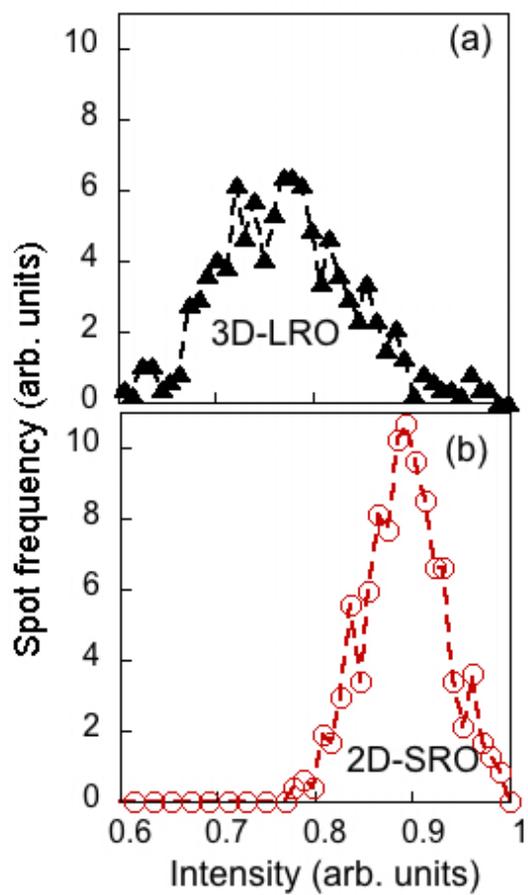

**Figure 3. (color online)** The intensities distribution of the 3D-LRO reflection intensities (filled triangles) (panel **a)** and 2D-SRO (empty circles) (panel **b)** for the 120 x 70 μm$^2$ sample surface region is shown.

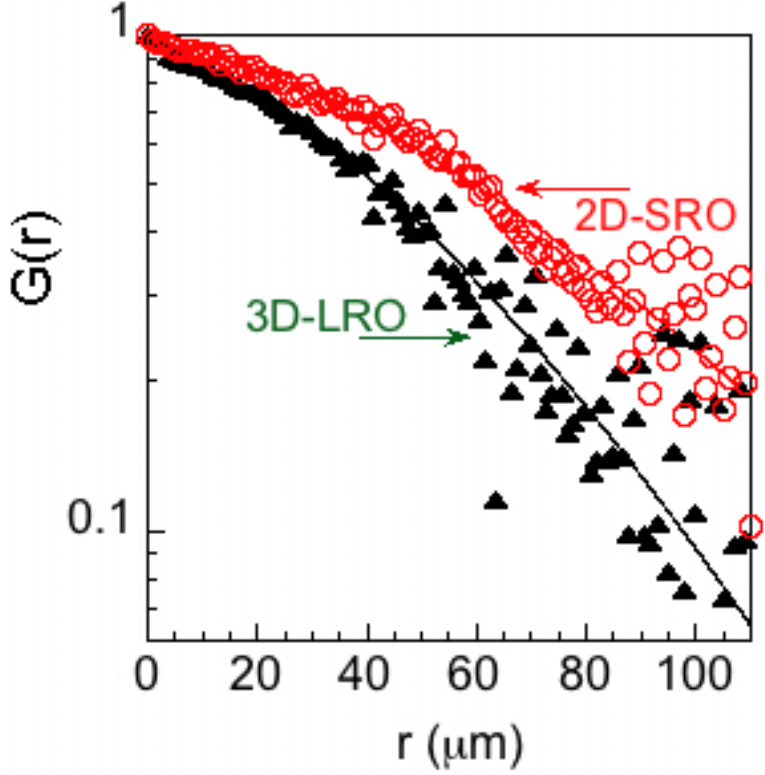

**Figure 4. (color online)** The spatial correlation function, G(r) of the x-ray diffraction intensities of the 3D-LRO superstructure reflections (filled triangles) and of the x-ray 2D-SRO reflections (empty circles) recorded by the micro XRD mapping at the spots on the surface of the sample. The solid lines show the fitting with the curves $G(r) = \langle G(\vec{r}) \rangle \propto A \exp[-r/\lambda]^{1.5}$ with correlation length $\lambda = 47 \mu m$ and $78 \mu m$ for the 3D-LRO and the 2D-SRO respectively.

The spatial correlation function G(r) calculated following Ref. [15] is plotted in Fig. 3c. The results show that the spatial correlation function can be described by compressed-exponential $G(r) = \langle G(\vec{r}) \rangle \propto A \exp[-r/\lambda]^{\beta}$ with the same exponent $\beta = 1.5 \pm 0.3$ and with a cut-off in the range of 30 μm to 80 μm . These spatial correlation lengths between corrugated domains are much larger than the domain size measured by the planar diffraction coherence lengths $\xi_b^{2D} = 82$ $nm$; $\xi_b^{3D} = 190$ $nm$;. The compressed exponential relaxation has been observed for a variety of soft matter systems undergoing 'jamming' transitions; -like collective dynamics with $\beta > 1$ [50] instead of liquid-like fluctuations with $\beta < 1$. It should be recalled that an exponent 1.5 has been reported for nematic liquid crystal on a surface [51] and for granular matter [52].

We have shown that the mixed k-space and real space visualization of the Bi2212 lattice shed new light on multiscale complexity in the HTS. These results contribute to the debate on rotational symmetry breaking observed by experimental methods measuring the response of the average structure over micron-size probed domains [53-58]. The large space variability in the amplitude of the lattice supermodulation underlines the high level of lattice disorder implying a disorder in the electronic structure. In particular small partial gaps due to the band folding determined by the striped lattice structure (superstripes) [8] could be masked by the spatial fluctuations of the lattice corrugation that breaks the translational symmetry. In fact a variation of 40 % in the amplitude of the structural one-dimensional modulation should make a fuzzy landscape of partial gaps and complex multiple Fermi surfaces. The left-handed distribution of superstructure fluctuation amplitude gives a chiral symmetry at the mcron-scale correlated with the breaking of the rotational symmetry of the rhombic $Q_2$ distortions of the $CuO_4$ platelets already observed by EXAFS and resonant diffraction [6-11,40]. The misift strain between the superconductor and the spacer layers [18-21] is proposed to be the external field that breaks the symmetry between the right-handed and the left-handed orientation of the $Q_2$ distortions [12,13]. The present data support the multiscale breaking of rotational symmetry at the microscale as well as at the previously observed atomic scale. Moreover these results introduces a new scenario of percolating microscopic superconducting domains in a preferential direction in a typical high temperature superconductor.

**Acknowledgements.** The authors are grateful to Ginestra Bianconi for discussions suggestions, and data analysis by using methods of statistical physics. We thank Manfred Burghammer and the ID13 beamline staff at ESRF for experimental help.

**References.**

[1] E. Dagotto, Science **309**, 257 (2005). http://dx.doi.org/10.1126/science.1107559
[2] A. R. Bishop, *Journal of Physics: Conference Series* **108**, 012027 (2008). http://dx.doi.org/10.1088/1742-6596/108/1/012027
[3] K. A. Müller, Journal of Physics: Condensed Matter **19**, 251002 (2007)
[4] A. Bianconi, *Physica C* **235-240**, 269-272 (1994); http://dx.doi.org/10.1016/0921-4534(94)91366-8 A. Bianconi, D. Di Castro, G. Bianconi, A. Pifferi, N. L. Saini, F. C. Chou, D. C. Johnston and M. Colapietro *Physica C* **341-348,** 1719 (2000). http://dx.doi.org/10.1016/S0921-4534(00)00950-3
[5] K. I. Kugel, A. L. Rakhmanov, A. O. Sboychakov, N. Poccia, and A. Bianconi, Phys. Rev. B **78**, 165124 (2008). http://dx.doi.org/10.1103/PhysRevB.78.165124
[6] A. Bianconi *Solid State Commun*. **91**, 1 (1994); http://dx.doi.org/10.1016/0038-1098(94)90831-1 A. Bianconi and M. Missori *Solid State Commun*. **91**, 287 (1994). http://dx.doi.org/10.1016/0038-1098(94)90304-2
[7] D. Innocenti, N. Poccia, A. Ricci, A. Valletta, S. Caprara, A. Perali, and A. Bianconi, Phys. Rev. B **82**, 184528 (2010). http://dx.doi.org/10.1103/PhysRevB.82.184528
[8] A. Bianconi, Intern. Jour. of Mod. Phys. B **14**, 3289 (2000); N. L. Saini and A. Bianconi, International Journal of Modern Physics B **14**, 3649 (2000).. http://dx.doi.org/10.1142/S0217979200004179
[9] A. Bianconi, A. Congiu Castellano, M. De Santis, P. Rudolf, P. Lagarde, A. M. Flank, and A. Marcelli *Solid State Commun*. **63**, 1009 (1987).
[10] A. Lanzara, P. V. Bogdanov, X. J. Zhou, S. A. Kellar, D. L. Feng, E. D. Lu, T. Yoshida, H. Eisaki, A. Fujimori, K. Kishio, J.-I. Shimoyama, T. Noda, S. Uchida, Z. Hussain and Z.-X. Shen Nature **412**, 510 (2001). http://dx.doi.org/10.1038/35087518
[11] T. Cuk, F. Baumberger, D. H. Lu, N. Ingle, X. J. Zhou, H. Eisaki, N. Kaneko, Z. Hussain, T. P. Devereaux, N. Nagaosa, and Z.-X. Shen Phys. Rev. Lett. 93, 117003 (2004). http://link.aps.org/abstract/PRL/v93/e117003
[12] Y. Seino, A. Kotani and A. Bianconi: J. Phys. Soc. Jpn. 59, 815 (1990). http://jpsj.ipap.jp/link?JPSJ/59/815
[13] A. Bianconi, N. L. Saini, A. Lanzara, M. Missori, T. Rossetti, H. Oyanagi, H. Yamaguchi, K. Oka, and T. Ito, Phys. Rev. Lett. **76**, 3412 (1996). http://dx.doi.org/10.1103/PhysRevLett.76.3412
[14] J. D. Jorgensen, B. Dabrowski, Shiyou Pei, D. G. Hinks, and L. Soderholm, B. Morosin, J. E. Schirber, E. L. Venturini, and D. S. Ginley Physical Review B 38, 11337 (1988). http://dx.doi.org/10.1103/PhysRevB.38.11337
[15] M. Fratini, N. Poccia, A. Ricci, G. Campi, M. Burghammer, G. Aeppli, and A. Bianconi, Nature **466**, 841 (2010). http://dx.doi.org/10.1038/nature09260
[16] N. Poccia, M. Fratini, A. Ricci, G. Campi, L. Barba, A. Vittorini-Orgeas, G. Bianconi, G. Aeppli and A. Bianconi Nature Materials, published online: 14 august 2011 | doi: 10.1038/nmat3088
[17] A. Bianconi *Solid State Communications* 89, 933 (1994). http://dx.doi.org/10.1016/0038-1098(94)90354-9
[18] A. Bianconi, S. Agrestini, G. Bianconi, D. Castro, and N. Saini, in Stripes and Related Phenomena, edited by A. Bianconi and N. L. Saini (Springer US, Boston, 2002), vol. 8 of Selected Topics in Superconductivity, chap. 2, pp. 9-25.
[19] A. Bianconi, N. L. Saini, S. Agrestini, D. Di Castro, and G. Bianconi, International Journal of Modern Physics B **14,** 3342 (2000). http://dx.doi.org/doi:10.1142/S0217979200003812
[20] A. Bianconi, D. Di Castro, N. L. Saini, and G. Bianconi, in *Phase Transitions and Self-Organization in Electronic and Molecular Network*s, edited by M. F. Thorpe and J. C. Phillips

(Kluwer Academic Publishers, Boston, 2002), Fundamental Materials Research, chap. 24, pag. 375. http://dx.doi.org/10.1007/0-306-47113-2_24 http://arxiv.org/abs/1107.4858

[21] A. Bianconi, G. Bianconi, S. Caprara, D. D. Castro, H. Oyanagi, and N. L. Saini, *Journal of Physics: Condensed Matter* **12**, 10655 (2000). http://dx.doi.org/10.1016/S0921-4534(00)00950-3

[22] M Fratini, N Poccia, and A Bianconi Journal of Physics: Conference Series, 108, 012036 (2008). http://dx.doi.org/10.1088/1742-6596/108/1/012036 arXiv:0812.136

[23] A. Ricci, N. Poccia, B. Joseph, L. Barba, G. Arrighetti, G. Ciasca, J. Q. Yan, R. W. McCallum, T. A. Lograsso, N. D. Zhigadlo, J. Karpinski, and A. Bianconi[1], Phys. Rev. B **82**, 144507 (2010). http://dx.doi.org/10.1103/PhysRevB.82.144507

[24] A. Ricci, N. Poccia, B. Joseph, G. Arrighetti, L. Barba, J. Plaisier, G. Campi, Y. Mizuguchi, H. Takeya, Y. Takano, N. L. Saini and A. Bianconi Superconductor Science and Technology **24**, 082002 (2011). http://iopscience.iop.org/0953-2048/24/8/082002/

[25] A. Ricci, N. Poccia, G. Campi, B. Joseph, G. Arrighetti, L. Barba, M. Reynolds, M. Burghammer, H. Takeya, Y. Mizuguchi, Y. Takano, M. Colapietro, N. L. Saini and A. Bianconi, preprint arXiv 1107.0412 (2011). http://arxiv.org/abs/1107.0412

[26] A. Yamamoto, M. Onoda, E. T. Muromachi, F. Izumi, T. Ishigaki, and H. Asano, Phys. Rev. B **42**, 4228 (1990). http://dx.doi.org/10.1103/PhysRevB.42.4228

[27] X. B. Kan and S. C. Moss, Acta Crystallographica Section B **48**, 122 (1992). http://dx.doi.org/10.1107/S0108768191011333

[28] A. A. Levin, Y. I. Smolin, and Y. F. Shepelev, Journal of Physics: Condensed Matter **6**, 3539 (1994). http://dx.doi.org/10.1088/0953-8984/6/19/009

[29] D. Grebille, H. Leligny, and O. Pérez, Phys. Rev. B **64**, 106501 (2001). http://dx.doi.org/10.1103/PhysRevB.64.106501

[30] A. Bianconi, M. Lusignoli, N. L. Saini, P. Bordet, and P. G. Radaelli, Phys. Rev. B **54**, 4310 (1996). http://dx.doi.org/10.1103/PhysRevB.54.4310

[31] A. Bianconi, N. L. Saini, T. Rossetti, A. Lanzara, A. Perali, M. Missori, H. Oyanagi, H. Yamaguchi, Y. Nishihara, and D. H. Ha, Physical Review B **54,** 12018 (1996).

[32] A. Perali, A. Bianconi, A. Lanzara, and N. L. Saini, Solid State Communications **100**, 181 (1996). http://dx.doi.org/10.1016/0038-1098(96)00373-0

[33] A. Bianconi, A. Valletta, A. Perali, and N. L. Saini, Solid State Communications **102**, 369 (1997) http://dx.doi.org/10.1016/S0038-1098(97)00011-2

[34] A. Bianconi, A. Valletta, A. Perali, and N. L. Saini, Physica C: Superconductivity **296**, 269 (1998). http://dx.doi.org/10.1016/S0921-4534(97)01825-X

[35] J. A. Slezak, J. Lee, M. Wang, K. McElroy, K. Fujita, B. M. Andersen, P. J. Hirschfeld, H. Eisaki, S. Uchida, and J. C. Davis, Proc. Nat. Acad. Sciences **105**, 3203 (2008). http://dx.doi.org/10.1073/pnas.0706795105

[36] Y. He, S. Graser, P. J. Hirschfeld, and H. P. Cheng, Phys. Rev. B **77**, 220507 (2008). http://dx.doi.org/10.1103/PhysRevB.77.220507

[37] K. Foyevtsova, H. C. Kandpal, H. O. Jeschke, S. Graser, H. P. Cheng, R. Valent'i, and P. J. Hirschfeld, Phys. Rev. B **82**, 054514 (2010). http://dx.doi.org/10.1103/PhysRevB.82.054514.

[38] A. Saxena, Y. Wu, T. Lookman, S. Shenoy, and A. Bishop, Physica A: Statistical Mechanics and its Applications **239**, 18 (1997). http://dx.doi.org/10.1103/PhysRevLett.91.057004

[39] J. X. Zhu, K. H. Ahn, Z. Nussinov, T. Lookman, A. V. Balatsky, and A. R. Bishop, *Phys. Rev. Lett*. **91**, 057004 (2003).

[40] N. Jenkins, Y. Fasano, C. Berthod, I. Maggio-Aprile, A. Piriou, E. Giannini, B. W. Hoogenboom, C. Hess, T. Cren, and Ø. Fischer, Phys. Rev. Lett. **103**, 227001 (2009). http://dx.doi.org/10.1103/PhysRevLett.103.227001

[41] A. Piriou, N. Jenkins, C. Berthod, I. Maggio-Aprile, and Ø. Fischer, Nature Communications **2**, 221 (2011). http://dx.doi.org/10.1038/ncomms1229

[42] C. V. Parker, P. Aynajian, E. H. da Silva Neto, A. Pushp, S. Ono, J. Wen, Z. Xu, G. Gu, and

A. Yazdani, Nature **468**, 677 (2010). http://dx.doi.org/10.1038/nature09597

[43] M. J. Lawler, K. Fujita, J. Lee, A. R. Schmidt, Y. Kohsaka, C. K. Kim, H. Eisaki, S. Uchida, J. C. Davis, J. P. Sethna, and Eun-Ah Kim Nature **466**, 347 (2010). http://dx.doi.org/10.1038/nature09169

[44] N. L. Saini, J. Avila, A. Bianconi, A. Lanzara, M. C. Asensio, S. Tajima, G. D. Gu, and N. Koshizuka, Phys. Rev. Lett. **79**, 3467 (1997). http://dx.doi.org/10.1103/PhysRevLett.79.3467

[45] I. Bdikin, A. N. Maljuk, b, A. B. Kulakov, C. T. Lin, P. Kumar, B. Kumar, G. C. Trigunayat and G. A. Emel'chenko Physica C: Superconductivity **383**, 431 (2003) http://dx.doi.org/10.1016/S0921-4534(02)01792-6

[46] A. Maljuk, B. Liang, C. T. Lin, and G. A. Emelchenko, Physica C: Superconductivity **355**, 140 (2001). http://dx.doi.org/10.1016/S0921-4534(00)01769-X

[47] T. Watanabe, T. Fujii, and A. Matsuda, Recent Res. Devel. Physics **5**, 51 (2004). arxiv:cond-mat/0401448

[48] N. L. Saini, H. Oyanagi, M. Molle, K. Garg, C. Kim, and A. Bianconi, Journal of Physics and Chemistry of Solids **65**, 1439 (2004). http://dx.doi.org/10.1016/j.jpcs.2003.12.011

[49] N. L. Saini, A. Lanzara, A. Bianconi, and H. Oyanagi, The European Phys. Jour. B - Condensed Matter and Complex Systems **18**, 257 (2000). http://dx.doi.org/10.1007/s100510070056

[50] O. G. Shpyrko, E. D. Isaacs, J. M. Logan, Y. Feng, G. Aeppli, R. Jaramillo, H. C. Kim, T. F. Rosenbaum, P. Zschack, M. Sprung, S. Narayanan and A. R. Sandy Nature **447**, 68 (2007). http://dx.doi.org/10.1038/nature05776

[51] M. Nespoulous, C. Blanc, and M. Nobili, Phys. Rev. Lett. 104, 097801 (2010), http://dx.doi.org/10.1103/PhysRevLett.104.097801

[52] P. Wang, C. Song, Y. Jin, K. Wang, and H. A. Makse, Journal of Statistical Mechanics: Theory and Experiment **2010**, P12005 (2010). http://dx.doi.org/10.1088/1742-5468/2010/12/P12005

[53] A. Kaminski, S. Rosenkranz, H. M. Fretwell, J. C. Campuzano, Z. Li, H. Raffy, W. G. Cullen, H. You, C. G. Olson, C. M. Varma, et al., Nature **416**, 610 (2002). http://dx.doi.org/10.1103/PhysRevLett.92.207001

[54] S. V. Borisenko, A. A. Kordyuk, A. Koitzsch, T. K. Kim, K. A. Nenkov, M. Knupfer, J. Fink, C. Grazioli, S. Turchini, and H. Berger, Phys. Rev. Lett. **92**, 207001 (2004). http://dx.doi.org/10.1103/PhysRevLett.92.207001

[55] M. Kubota, K. Ono, Y. Oohara, and H. Eisaki, Jour. of the Phys. Soc. Japan **75**, 053706 (2006) http://dx.doi.org/10.1143/JPSJ.75.053706

[56] V. Arpiainen, A. Bansil and M. Lindroos, Phys. Rev. Lett. **103**, 067005 (2009). http://dx.doi.org/10.1103/PhysRevLett.103.067005

[57] M. R. Norman, A. Kaminski, S. Rosenkranz, and J. C. Campuzano, Phys. Rev. Lett.**105,** 189701 (2010). http://dx.doi.org/10.1103/PhysRevLett.105.189701.103

[58] R. Daou, J. Chang, D. LeBoeuf, O. Cyr-Choiniere, F. Laliberte, N. Doiron-Leyraud, B. J. Ramshaw, R. Liang, D. A. Bonn, W. N. Hardy and Louis Taillefer Nature **463**, 519 (2010). http://dx.doi.org/10.1038/nature08716.